\documentclass[twocolumn,english,superscriptaddress]{revtex4}
\usepackage[T1]{fontenc}
\usepackage[latin1]{inputenc}
\usepackage{graphicx}
\usepackage{amssymb}

\makeatletter

\newcommand{\boldsymbol}[1]{\mbox{\boldmath $#1$}}

\providecommand{\tabularnewline}{\\}

\usepackage{graphicx}
\usepackage{graphicx}
\usepackage{graphicx}
\usepackage{graphicx}

\usepackage{epsf}

\usepackage{babel}
\makeatother
\begin{document}

\title{Power Law Blinking Quantum Dots: Stochastic and Physical Models}

\author{Gennady Margolin}

\affiliation{Department of Chemistry and Biochemistry, Notre Dame University,
Notre Dame, IN 46556}

\author{Vladimir Protasenko}

\affiliation{Department of Chemistry and Biochemistry, Notre Dame University,
Notre Dame, IN 46556}

\author{Masaru Kuno}

\affiliation{Department of Chemistry and Biochemistry, Notre Dame University,
Notre Dame, IN 46556}

\author{Eli Barkai}

\affiliation{Department of Chemistry and Biochemistry, Notre Dame University,
Notre Dame, IN 46556}

\affiliation{Department of Physics, Bar Ilan University, Ramat Gan, Israel 52900}

\date{\today{}}

\begin{abstract}
We quantify nonergodic and aging behaviors of nanocrystals (or quantum
dots) based on stochastic model. Ergodicity breaking is characterized
based on time average intensity and time average correlation function,
which remain random even in the limit of long measurement time. We
argue that certain aspects of nonergodicity can be explained based
on a modification of Onsager's diffusion model of an ion pair escaping
neutralization. We explain how diffusion models generate nonergodic
behavior, namely a simple mechanism is responsible for the breakdown
of the standard assumption of statistical mechanics. Data analysis
shows that distributions of on and off intervals in the nanocrystal
blinking are almost identical, $\psi_{\pm}(\tau)\propto A_{\pm}\tau^{-(1+\alpha_{\pm})}$
with $A_{+}\approx A_{-}$ and $\alpha_{+}\approx\alpha_{-}=\alpha$
and $\alpha\approx0.8$. The latter exponent indicates that a simple
diffusion model with $\alpha=0.5$ neglecting the electron-hole Coulomb
interaction and/or tunneling, is not sufficient. 
\end{abstract}
\maketitle

\section{Introduction}

Single quantum dots when interacting with a continuous wave laser
field blink: at random times the dot turns from a state on, in which
many photons are emitted to a state off in which no photons are emitted.
While stochastic intensity trails are found today in a vast number
of single molecule experiments, the dots exhibit statistical behavior
which seems unique. In particular, the dots exhibit power law statistics,
aging, and ergodicity breaking. While our understanding of the Physical
origin of the blinking behavior of the dots is not complete, several
physical pictures have emerged in recent years, which explain the
blinking in terms of simple Physics. Here we will review a diffusion
model which might explain some of the observations made so far. Then
we analyze the stochastic properties of the dots, using a stochastic
approach. In particular we review the behaviors of the time and ensemble
average intensity correlation functions. Usually it is assumed that
these two objects are identical in the limit of long times, however
this is not the case for the dots.

\section{Physical Models\label{sec:Physical-Meaning}}

A typical fluorescence intensity trace of a CdSe quantum dot, or nanocrystal
(NC), overcoated with ZnS (in short, CdSe-ZnS NC) under continuous
laser illumination is shown in Fig.%
\begin{figure}
\includegraphics[%
  width=1.0\columnwidth,
  keepaspectratio]{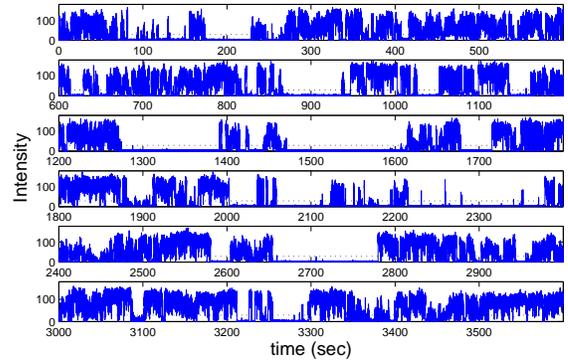}

\caption{\label{cap:Kuno1}Intensity fluctuations in a CdSe-ZnS NC under continuous
laser illumination at room temperature. Dotted horizontal line was
selected as a threshold to divide off and on states.}
\end{figure}
\ref{cap:Kuno1}. From this Figure we learn, that roughly, the intensity
jumps between two states - on and off. Some of the deviations from
this digital behavior can be attributed to fluctuating non-radiative
decay channels due to coupling to the environment, and also to time
binning procedure \cite{Schlegel02,Fisher04,ChungBawendi04}, and
see also \cite{Verberk02}. Data analysis of such time trace is many
times based on distribution of on and off times. Defining a threshold
above which the NC is considered in state on and under which it is
in state off, one can extract the probability density functions $\psi_{+}(\tau)$
of on and $\psi_{-}(\tau)$ of off times. Surprisingly these show
a power-law decay $\psi_{\pm}(\tau)\propto\tau^{-1-\alpha_{\pm}}$,
as shown in Fig.%
\begin{figure}
\includegraphics[%
  width=1.0\columnwidth,
  keepaspectratio]{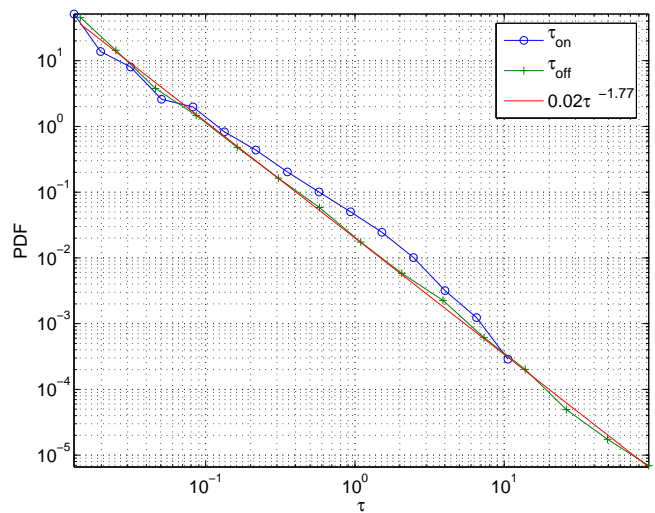}

\caption{\label{cap:Kuno2}Distributions $\psi_{\pm}(\tau)$ of on and off
times for the NC, whose intensity trajectory is shown in Fig. \ref{cap:Kuno1}.
The straight line is the fit to the off time distribution.}
\end{figure}
\ref{cap:Kuno2}. A summary of different experimental exponents is
presented in Table%
\begin{table*}
\begin{tabular}{|c|c|c|c|c|c|c|c|}
\hline 
\textbf{Group}&
 \textbf{Material}&
 \textbf{No.}&
 \textbf{Radii, nm}&
 \textbf{Temp., K}&
 \textbf{\footnotesize Laser Intensity,} \textbf{$\frac{\textrm{kW}}{\textrm{cm}^{2}}$}&
 \boldsymbol{\alpha_{+}}&
 \boldsymbol{\alpha_{-}}\tabularnewline
\hline
Verberk et al. \cite{Verberk02}&
CdS&
1&
2.5&
1.2&
&
$e^{-at}$&
0.65(0.2)\tabularnewline
\hline
Brokmann et al. \cite{Brokmann03}&
 CdSe-ZnS&
 215&
&
 300&
&
 0.58(0.17)&
 0.48(0.15)\tabularnewline
\hline
Shimizu et al. \cite{Shimizu01}&
 CdSe-ZnS, CdSe, CdTe&
 >200&
 1.5, 2.5&
 300, 10&
 0.1-0.7&
 0.5(0.1), cutoff&
 0.5(0.1)\tabularnewline
\hline
Kuno et al. \cite{Kuno00}&
 CdSe-ZnS&
 $\sim200$&
 1.7-2.9&
 300-394&
 0.24-2.4&
&
 0.5-0.75\tabularnewline
\hline
Kuno et al. \cite{Kuno01}&
 CdSe-ZnS&
 >300&
 1.7-2.7&
 300&
 0.1-100&
 0.9(0.05)&
 0.54(0.03)\tabularnewline
\hline
Kuno et al. \cite{Kuno01_2}&
 InP&
 $\sim30$&
 1.5&
 300&
 0.24&
 1.0(0.2)&
 0.5(0.1)\tabularnewline
\hline
Cichos et al. \cite{Cichos04}&
 Si&
&
&
 &
 1.8, 6.5&
 1.2(0.1)&
 0.3, 0.7\tabularnewline
\hline
Hohng and Ha \cite{HohngHa04}&
 CdSe-ZnS&
 $\sim1000$&
&
 &
&
&
 0.94-1.10\tabularnewline
\hline
M\"{u}ller et al. \cite{Muller04}&
 CdSe-ZnS&
&
 4.4 (core)&
 300&
 0.025&
 0.55&
 0.05, 0.25\tabularnewline
\hline
van Sark et al. \cite{vanSark02}&
 CdSe-ZnS&
 41&
 $\sim3.7$&
 300&
 20&
 $\sim1.2$, $\sim0.7$&
 $\sim0.2$, $\sim0.4$\tabularnewline
\hline
Kobitski et al. \cite{Kobitski04}&
 CdSe&
&
 3.6&
 &
 0.04-0.38&
 0.97-0.66&
 0.42-0.64 \tabularnewline
\hline
\end{tabular}

\caption{\label{cap:Summary-on-off}Summary of experimental exponents for
on ($\alpha_{+}$) and off ($\alpha_{-}$) time distributions for
various single NCs under different experimental conditions. Notice
that Verberk et al. use uncapped NCs, while other measurement consider
capped NCs, hence exponential distribution on times is found only
for uncapped dots. Hohng and Ha used CdSe-ZnS NCs coated with streptavidin
which might alter the exponent $\alpha_{-}$.}
\end{table*}
\ref{cap:Summary-on-off}, indicating such a power-law decay in most
cases. In some cases $\alpha_{+}\approx\alpha_{-}$ and the exponents
are close to 1/2. In particular, Brokmann et al. \cite{Brokmann03}
measured 215 CdSe-ZnS NCs and found that all are statistically identical
with $\alpha_{+}=0.58\pm0.17$, $\alpha_{-}\approx0.48\pm0.15$ so
that $\alpha_{+}\approx\alpha_{-}\approx0.5$. Note that most of the
uncertainty in the values of the exponents can be attributed simply
to statistical limitations of data analysis \cite{Issac05} (see also
Section \ref{sec:Experimental-evidence} below). Shimizu et al. \cite{Shimizu01}
found that in the limit of low temperature and weak laser fields $\alpha_{+}\approx\alpha_{-}\approx0.5$.
The fact that in many cases $\alpha_{\pm}<1$, leads to interesting
statistical behavior, for example ergodicity breaking, and aging.
We will discuss these behaviors in Sec. III. A physical model for
blinking was suggested by Efros and Rosen \cite{EfrosRosen97}. Briefly
the on and off periods correspond to neutral and charged NCs respectively.
Thus the on/off trace teaches us something on elementary charging
mechanism of the dot. The difficulty is to explain the power law distributions
of on and off times, or in other words why should the time the charge
occupies the NCs follow power law behavior? 

Two types of models were suggested, a diffusion approach and a random
trap model. The measurements of Dahan's and Bawendi's groups \cite{Brokmann03,Shimizu01},
which show the universal power law $\alpha_{\pm}=0.5$, are consistent
with the diffusion model (see details below). The fact that all dots
are found to be similar \cite{Brokmann03} seem not consistent with
models of quenched disorder \cite{Verberk02,Kuno01,Kuno03} since
these support the idea of a distribution of $\alpha_{\pm}$. However,
some experiments show deviations from the $\alpha_{+}\approx\alpha_{-}\approx0.5$
and might support the distribution of $\alpha_{\pm}$. It is possible
that preparation methods and environments lead to different mechanisms
of power law blinking, and different exponents \cite{Issac05}. More
experimental work in this direction is needed, in particular, experimentalists
still have to investigate the distribution of $\alpha_{\pm}$, and
show whether and under what conditions are all the dots statistically
identical. Below we discuss the diffusion model; different aspects
of the tunneling and trapping model can be found in \cite{Verberk02,Kuno03,Issac05}. 

As discussed at length by Shimizu et al. \cite{Shimizu01}, the on
time distributions show temperature and laser power dependencies,
e.g. exponential cutoffs of power law behavior. Although no direct
observations of cutoffs in the off time distribution was reported,
ensemble measurements by Chung and Bawendi \cite{ChungBawendi04}
demonstrate that there should be such a cutoff as well, but at times
of the order of tens of minutes to hours. Our analysis here, employing
the power law decaying distributions, is of course applicable in time
windows where power law statistics holds.

\subsection{Diffusion model}

We note that the simplest diffusion controlled chemical reaction $A+B\rightleftharpoons AB$,
where $A$ is fixed in space, can be used to explain some of the observed
behavior on the uncapped NCs. As shown by the group of Orrit \cite{Verberk02}
such dots exhibit exponential distribution of \emph{on} times and
power law distribution of \emph{off} times. The \emph{on} times follow
standard exponential kinetics corresponding to an ionization of a
neutral NC (denoted as \emph{AB}). A model for this exponential behavior
was given already in \cite{EfrosRosen97}. Clearly the experiments
of the group of Orrit, show that the capping plays an important part
in the the blinking, since capped NCs exhibit power law behavior both
for the on and off times. We will return to capped dots later. 

Once the uncapped NC is ionized ($A+B$ state) we assume the ejected
charge carrier exhibits a random walk on the surface of the NC or
in the bulk. This part of the problem is similar to Onsager's classical
problem of an ion pair escaping neutralization (see e.g., \cite{HongNoolandi78,SanoTachiya79}).
The survival probability in the \emph{off} state for time \emph{t},
$S_{-}(t)$ is related to the \emph{off} time distribution via $S_{-}(t)=1-\int_{0}^{t}\psi_{-}(\tau)d\tau$,
or \begin{equation}
\psi_{-}(t)=-\frac{dS_{-}(t)}{dt}.\end{equation}
 It is well known that in three dimensions survival probability decays
like $t^{-1/2}$, the exponent 1/2 is close to the exponent often
measured in the experiments. In infinite domain the decay is not to
zero, but the 1/2 appears in many situations, for finite and infinite
systems, in completely and partially diffusion controlled recombination,
in different dimensions, and can govern the leading behavior of the
survival probability for orders of magnitude in time \cite{SanoTachiya79,NadlerStein91,NadlerStein96}.
In this picture the exponent $1/2$ does not depend on temperature,
similar to what is observed in experiment. We note that it is possible
that instead of the charge carrier executing the random walk, diffusing
lattice defects which serve as a trap for charge carrier are responsible
for the blinking behavior of the NCs. 

A long time ago, Hong, Noolandi and Street \cite{HongNoolandiStreet81}
investigated geminate electron-hole recombination in amorphous semiconductors.
In their model they included the effects of tunneling, Coulomb interaction,
and diffusion. Combination of tunneling and diffusion leads to a $S(t)\propto t^{-1/2}$
behavior. However, when the Coulomb interactions are included in the
theory, deviations from the universal $t^{-1/2}$ law, are observed.
For example in the the analysis of photoluminescence decay in amorphous
Si:H, as a function of temperature. 

Coulomb interaction between the charged NCs and the ejected electron
seems to be an important factor in the Physics of NCs. The Onsager
radius is a measure of the strength of the interaction \begin{equation}
r_{Ons}=\frac{e^{2}}{k_{b}T\epsilon}.\end{equation}
 Krauss and Brus \cite{KraussBrus99} measured the dielectric constant
of CdSe dots, and found the value of $8$. Hence, at room temperature
we find $r_{Ons}\simeq70\textrm{Å}$ (however, note that the dielectric
constant of the matrix is not identical to that of the dot). Since
the length scale of the dots is of the order of a few nanometers,
the Coulomb interaction seems an important ingredient of the problem.
This according to the theory in \cite{HongNoolandiStreet81} is an
indication of possible deviations from the universal $1/2$ power
law behavior. It is also an indication that an ejected electron is
likely to return to the dot and not escape to the bulk (since the
force is attractive). In contrast, if the Onsager radius is small,
an ejected electron would most likely escape to the bulk, leaving
the dot in state off forever (i.e. Polya theorem in three dimensions).
Unfortunately, currently there is not sufficient experimental data
to determine in more qualitative ways if, Onsager type of model can
be used to explain the observed data. As in standard geminate recombination
processes, the dependence of blinking on temperature, dielectric constant
of the dot and of the matrix \cite{Issac05}, and on external driving
field, might yield more microscopical information on the precise physical
mechanism of the fascinating blinking behavior. 

One of the possible physical pictures explaining blinking of capped
NCs can be based on diffusion process, using a variation of a three
state model of Verberk et al. \cite{Verberk02}. As mentioned above,
for this case power law distribution of \emph{on} and \emph{off} times
are observed. In particular, neutral capped NC will correspond to
state \emph{on} (as for uncapped NCs). However, capped NC can remain
\emph{on} even in the ionized state - see Fig.%
\begin{figure}
\includegraphics[%
  width=1.0\columnwidth,
  keepaspectratio]{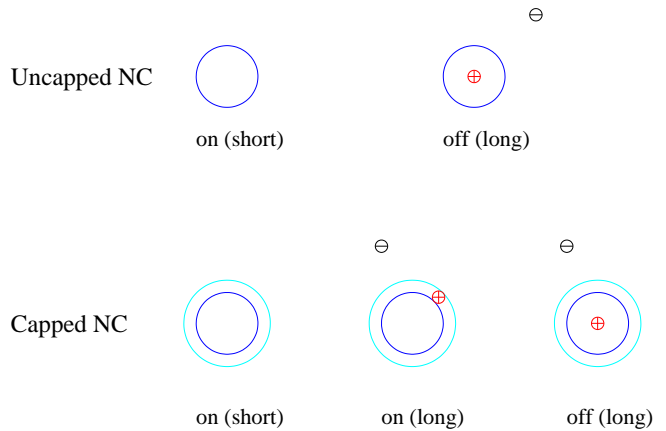}

\caption{\label{cap:On-and-off}On and off states for NCs, following \cite{Verberk02}.}
\end{figure}
\ref{cap:On-and-off}. Verberk et al. assume that the ionized capped
NC can be found in two states: (i) the charge remaining in the NC
can be found in center of NC (possibly a de-localized state), (ii)
charge remaining in the NC can be trapped in vicinity of capping.
For case (i) the NC will be in state \emph{off}, for case (ii) the
NC will be in state \emph{on}. Depending on exact location of this
charge, the fluorescence intensity can vary. The main idea is that
the rate of Auger nonradiative recombination \cite{EfrosRosen97}
of consecutively formed electron-hole pairs will drop for case (ii)
but not for case (i). We note that capping may increase effective
radius of the NC, or provide trapping sites for the hole (e.g., recent
studies by Lifshitz et al. \cite{Lifshitz04} demonstrate that coating
of NCs creates trapping sites in the interface). Thus the \emph{off}
times occur when the NC is ionized and the hole is close to the center,
these \emph{off} times are slaved to the diffusion of the electron.
While \emph{on} times occur for both a neutral NC and for charged
NC with the charge in vicinity of capping, the latter \emph{on} times
are slaved to the diffusion of the electron. In the case of power
law \emph{off} time statistics this model predicts same power law
exponent for the \emph{on} times, because both of them are governed
by the return time of the ejected electron. 

Beyond nanocrystals, we note that fluorescence of single molecules
\cite{Haase04} and of nanoparticles diffusing through a laser focus
\cite{Zumofen04}, switching on and off of vibrational modes of a
molecule \cite{Bizzarri05}, opening-closing behavior of certain single
ion channels \cite{NadlerStein91,Goy02,Goy03}, motion of bacteria
\cite{Korobkova04}, deterministic diffusion in chaotic systems \cite{ZK},
the sign of magnetization of spin systems at criticality \cite{GL},
and others exhibit power law intermittency behavior \cite{MB_condmat05}.
More generally the time trace of the NCs is similar to the well known
L\'{e}vy walk model \cite{JBS02}. Hence the stochastic theory which
we consider in the following section is very general. In particular
we do not restrict our attention to the exponent 1/2, as there are
indications for other values of $\alpha$ between 0 and 1, and the
analysis hardly changes.

\section{Stochastic Model and Definitions\label{SecModel}}

\begin{figure}
\includegraphics[%
  width=1.0\columnwidth,
  keepaspectratio]{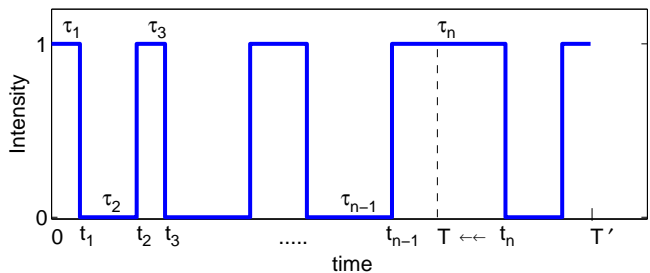}

\caption{\label{cap:figure-def}Schematic temporal evolution of the dichotomous
intensity process.}
\end{figure}

The random process considered in this manuscript, is shown in Fig.
\ref{cap:figure-def}. The intensity $I(t)$ jumps between two states
$I(t)=+1$ and $I(t)=0$. At start of the measurement $t=0$ the NC
is in state \emph{on}: $I(0)=1$. The sojourn time $\tau_{i}$ is
an \emph{off} time if $i$ is even, it is an \emph{on} time if $i$
is odd (see Fig. \ref{cap:figure-def}). The times $\tau_{i}$ for
odd {[}even{]} $i$, are drawn at random from the probability density
function (PDF) $\psi_{+}(t)$, $[\psi_{-}(t)]$, respectively. These
sojourn times are mutually independent, identically distributed random
variables. Times $t_{i}$ are cumulative times from the process starting
point at time zero till the end of the \emph{i}'th transition. Time
$T'$ on Fig. \ref{cap:figure-def} is the time of observation. We
denote the Laplace transform of $\psi_{\pm}(t)$ using \begin{equation}
\hat{\psi}_{\pm}(s)=\int_{0}^{\infty}\psi_{\pm}(t)e^{-st}\textrm{d}t.\end{equation}

In what follows we will investigate statistical properties of this
seemingly simple stochastic process. In particular we will investigate
the correlation function of this process. In experiment correlation
functions are used many times to characterize intensity trajectories.
The main advantage of the analysis of correlation functions, if compared
with PDFs of on and off times, is that in former case there is no
need to introduce the intensity cutoff. Correlations functions are
more general than on and off time distributions. Besides, correlation
functions exhibit aging, and ergodicity breaking, which are in our
opinion interesting.

We will consider several classes of on/off PDFs, and classify generic
behaviors based on the small $s$ expansion of $\psi_{\pm}\left(s\right)$.
We will consider:\\
 (i) \textbf{Case 1} PDFs with finite mean \emph{on} and \emph{off}
times, whose Laplace transform in the limit $s\rightarrow0$ satisfies:
\begin{equation}
\hat{\psi}_{\pm}(s)=1-s\tau_{\pm}+\cdots.\label{eq0m1}\end{equation}
 Here $\tau_{+}$ $(\tau_{-})$ is the average \emph{on} (\emph{off})
time. For example exponentially distributed \emph{on} and \emph{off}
times, \begin{equation}
\hat{\psi}_{\pm}(s)=\frac{1}{1+s\tau_{\pm}},\label{EqExp}\end{equation}
 belong to this class of PDFs.\\
 (ii) \textbf{Case 2} PDFs with infinite mean \emph{on} and \emph{off}
times, namely PDFs with power law behavior satisfying \begin{equation}
\psi_{\pm}\propto t^{-1-\alpha_{\pm}}\,\,\,\,\,\,\,\,\,\,\,\,\alpha_{-}<\alpha_{+}\leq1,\end{equation}
 in the limit of long times. The small $s$ behavior of these family
of functions satisfies \begin{equation}
\hat{\psi}_{\pm}(s)=1-A_{\pm}s^{\alpha_{\pm}}+\cdots\label{eqm0}\end{equation}
 where $A_{\pm}$ are parameters which have units of time$^{\alpha}$.
We will also consider cases where \emph{on} times have finite mean
$(\alpha_{+}=1)$ while the \emph{off} mean time diverges $(\alpha_{-}<1)$
since this situation describes behavior of uncapped NC \cite{Verberk02}
(see also \cite{Novikov03}). \\
 (iii) \textbf{Case 3} PDFs with infinite mean with $\alpha_{+}=\alpha_{-}=\alpha$\begin{equation}
\hat{\psi}_{\pm}(s)=1-A_{\pm}s^{\alpha}+\cdots\label{eqm0a}\end{equation}
 As mentioned Brokmann et al. \cite{Brokmann03} report that for CdSe
dots, $\alpha_{+}=0.58\pm0.17$, and $\alpha_{-}=0.48\pm0.15$, hence
within error of measurement, $\alpha\simeq0.5$. 

Standard theories of data analysis, usually use the ergodic hypothesis
and a time average of a process is replaced with an average over an
ensemble. The simplest time average in our case is the time average
intensity \begin{equation}
\overline{I}=\frac{\int_{0}^{T'}\,\, I(t)\textrm{d}t}{T'}.\end{equation}
 In the limit of long times and if ergodic assumption holds $\overline{I}=\langle I\rangle$,
where $\langle I\rangle$ is the ensemble average. As usual we may
generate many intensity trajectories one at a time, to obtain ensemble
averaged correlation function\begin{equation}
C(t,t')=\langle I(t)I(t+t')\rangle,\label{eq0DEF}\end{equation}
 and the normalized ensemble averaged correlation function\begin{equation}
g^{(2)}(t,t')\equiv\frac{\langle I(t)I(t+t')\rangle}{\left\langle I(t)\right\rangle \left\langle I(t+t')\right\rangle }=\frac{C(t,t')}{\left\langle I(t)\right\rangle \left\langle I(t+t')\right\rangle },\label{eq:g2def}\end{equation}
 From a single trajectory of $I(t)$, recorded in a time interval
$(0,T')$, we may construct the time average (TA) correlation function
\begin{equation}
C_{TA}(T',t')=\frac{\int_{0}^{T'-t'}I(t)I(t+t')\textrm{d}t}{T'-t'}.\label{eqTA}\end{equation}
 In single molecule experiments, the time averaged correlation function
is considered, not the ensemble average. However, it is many times
assumed that the ensemble average and the time average correlation
functions are identical. For nonergodic processes $C_{TA}(T',t')\ne C(t,t')$
even in the limit of large $t$ and $T'$. Moreover for nonergodic
processes, even in the limit of $T'\rightarrow\infty$, $C_{TA}(T',t')$
is a random function which varies from one sample of $I(t)$ to another.
The ensemble-averaged function $C(t,t')$ of the considered process
is non-stationary, i.e., it keeps its dependence on $t$ even when
$t\rightarrow\infty$. This is known as aging. It follows then from
Eq. (\ref{eqTA}) that $\left\langle C_{TA}(T',t')\right\rangle =\int_{0}^{T'-t'}C(t,t')dt/(T'-t')\neq C(t,t')$.

\section{Aging}

Consider the ensemble averaged correlation function $C(t,t')=\langle I(t+t')I(t)\rangle$.
For processes with finite microscopical time scale, which exhibit
stationary behavior, one has $C(t,t')=f(t')$. Namely the correlation
function does not depend on the observation time $t$. Aging means
that $C(t,t')$ depends on both $t$ and $t'$ even in the limit when
both are large \cite{Bouchaud92,Aquino04}. Simple aging behavior
means that at the scaling limit $C(t,t')=f(t'/t)$, which is indeed
the scaling in our Case 3; in Case 2 below we find such a scaling
for $g^{(2)}(t,t')$, while $C(t,t')$ will scale differently. Aging
and non-ergodicity are related. In our models, when single particle
trajectories turn non-ergodic, the ensemble average exhibit aging.
Both behaviors are related to the fact that there is no characteristic
time scale for the underlying process.

\subsection{Mean Intensity of \emph{on-off} process\label{SecInt}}

The ensemble averaged intensity $\langle I(t)\rangle$ for the process
switching between 1 and 0 and starting at 1 is now considered, which
will be used later. In Laplace $t\rightarrow s$ space it is easy
to show that \begin{equation}
\left\langle \hat{I}(s)\right\rangle =\frac{1-\hat{\psi}_{+}(s)}{s}\cdot\frac{1}{1-\hat{\psi}_{+}(s)\hat{\psi}_{-}(s)}.\label{EqMeanIntensity}\end{equation}
 The Laplace $s\rightarrow t$ inversion of Eq. (\ref{EqMeanIntensity})
yields the mean intensity $\langle I(t)\rangle$. Using small $s$
expansions of Eq. (\ref{EqMeanIntensity}), we find in the limit of
long times \begin{equation}
\langle I(t)\rangle\sim\left\{ \begin{array}{cc}
\frac{\tau_{+}}{\tau_{+}+\tau_{-}} & \mbox{case}\,\,\,\,1\\
\,\, & \,\,\\
\frac{A_{+}t^{\alpha_{-}-\alpha_{+}}}{A_{-}\Gamma\left(1+\alpha_{-}-\alpha_{+}\right)} & \mbox{case}\,\,\,\,2\\
\,\, & \,\,\\
\frac{A_{+}}{A_{+}+A_{-}} & \mbox{case}\,\,\,\,3.\end{array}\right.\label{eqIave}\end{equation}

If the \emph{on} times are exponential, as in Eq. (\ref{EqExp}) then
\begin{equation}
\left\langle \hat{I}(s)\right\rangle =\frac{\tau_{+}}{1+s\tau_{+}-\psi_{-}(s)}.\label{eq:Iexpon}\end{equation}
 This case corresponds to the behavior of the uncapped NCs. The expression
in Eq. (\ref{eq:Iexpon}), and more generally, the case $\alpha_{-}<\alpha_{+}=1$
leads for long time \emph{t} to \begin{equation}
\left\langle I(t)\right\rangle \sim\frac{\tau_{+}t^{\alpha_{-}-1}}{A_{-}\Gamma(\alpha_{-})}.\label{eqiii}\end{equation}
 For exponential \emph{on} and \emph{off} time distributions Eq. (\ref{EqExp}),
we obtain the exact solution \begin{equation}
\left\langle I(t)\right\rangle =\frac{\tau_{-}\exp\left[-t\left(\frac{1}{\tau_{-}}+\frac{1}{\tau_{+}}\right)\right]+\tau_{+}}{\tau_{-}+\tau_{+}}.\label{eqExp}\end{equation}

The average intensity does not yield direct evidence for aging, because
it depends only on one time variable, and one has to consider a correlation
function to explore aging in its usual meaning. 

\textbf{Remark} For the case $\alpha_{+}<\alpha_{-}<1$, corresponding
to a situation where \emph{on} times are in statistical sense much
longer then \emph{off} times, $\langle I(t)\rangle\sim1$.

\subsection{Aging Correlation Function of \emph{on}-\emph{off} process\label{SecAGE}}

The ensemble averaged correlation function $C(t,t')=\langle I(t)I(t+t')\rangle$
was calculated in \cite{MB_JCP04}. Contributions to the correlation
function arise only from trajectories with $I(t)=1$ and $I(t+t')=1$,
yielding\[
\hat{C}(t,u)=\frac{\hat{f}_{t}(u=0,+)-\hat{f}_{t}(u,+)}{u}\]
\begin{equation}
+\hat{f}_{t}(u,+)\frac{\hat{\psi}_{-}(u)\left[1-\hat{\psi}_{+}(u)\right]}{u\left[1-\hat{\psi}_{-}(u)\hat{\psi}_{+}(u)\right]},\label{eq:Corr}\end{equation}
 where $u$ is the Laplace conjugate of $t'$ and \begin{equation}
\hat{f}_{s}(u,+)=\frac{\hat{\psi}_{+}(s)-\hat{\psi}_{+}(u)}{(u-s)\left[1-\hat{\psi}_{+}(s)\hat{\psi}_{-}(s)\right]},\label{eq:f+}\end{equation}
 where $s$ is the Laplace conjugate of $t$. We note that $\hat{f}_{s}(u,+)$
is the double Laplace transform of the PDF of the so called forward
recurrence time. This means that after the aging of the process in
time interval $t$, the statistics of first jump event after time
$t$ will generally depend on the age $t$. However, a process is
said to exhibit aging, only if the statistics of this first jump depend
on $t$ even when this age is long. In particular if the microscopical
time scale of the problem is infinite, no matter how big is $t$ the
correlation function still depends on the age (see details below).
The first term in Eq. (\ref{eq:Corr}) is due to trajectories which
were in state on at time $t$ and did not make any transitions (i.e.
the concept of persistence), while the second term includes all the
contributions from the trajectories being in state on at time $t$
and making an even number of transitions \cite{MB_JCP04}.

\subsection{Case 1}

For case $1$ with finite $\tau_{+}$ and $\tau_{-}$, and in the
limit of long times $t$, we find \[
\lim_{t\rightarrow\infty}\hat{C}(t,u)=\]
\begin{equation}
\frac{1}{u}\frac{\tau_{+}}{\tau_{+}+\tau_{-}}\left\{ 1-\frac{\left[1-\hat{\psi}_{+}(u)\right]\left[1-\hat{\psi}_{-}(u)\right]}{\tau_{+}u\left[1-\hat{\psi}_{-}(u)\hat{\psi}_{+}(u)\right]}\right\} \label{eqVerO}\end{equation}
 This result was obtained by Verberk and Orrit \cite{Verberk03} and
it is seen that the correlation function depends asymptotically only
on $t'$ (since \emph{u} is Laplace pair of $t'$). Namely, when average
\emph{on} and \emph{off} times are finite the system does not exhibit
aging. If both $\psi_{+}(t)$ and $\psi_{-}(t)$ are exponential then
the \emph{exact} result is\[
\begin{array}{ccc}
C(t,t') & = & {\displaystyle \frac{\tau_{-}\exp\left[-t\left(\frac{1}{\tau_{-}}+\frac{1}{\tau_{+}}\right)\right]+\tau_{+}}{\tau_{-}+\tau_{+}}}\\
\\ & \times & {\displaystyle \frac{\tau_{-}\exp\left[-t'\left(\frac{1}{\tau_{-}}+\frac{1}{\tau_{+}}\right)\right]+\tau_{+}}{\tau_{-}+\tau_{+}}}\end{array}\]
 and $C(t,t')$ becomes independent of \emph{t} exponentially fast
as \emph{t} grows.

\subsection{Case 2}

We consider case $2$, however limit our discussion to the case $\alpha_{+}=1$
and $\alpha_{-}<1$. As mentioned this case corresponds to uncapped
NCs where \emph{on} times are exponentially distributed, while \emph{off}
times are described by power law statistics. Using the exact solution
Eq. (\ref{eq:Corr}) we find asymptotically, when both \emph{t} and
$t'$ are large: \begin{equation}
C(t,t')\sim\left(\frac{\tau_{+}}{A_{-}}\right)^{2}\frac{\left(tt'\right)^{\alpha_{-}-1}}{\Gamma^{2}\left(\alpha_{-}\right)}.\label{eqG1}\end{equation}
 Unlike case $1$ the correlation function approaches zero when $t\rightarrow\infty$,
since when $t$ is large we expect to find the process in state \emph{off}.
Using Eq. (\ref{eqiii}), the asymptotic behavior of the normalized
correlation function Eq. (\ref{eq:g2def}) is \begin{equation}
g^{(2)}(t,t')\sim\left(1+\frac{t}{t'}\right)^{1-\alpha_{-}}.\label{eq:case2g2asymp}\end{equation}
 We see that the correlation functions Eqs. (\ref{eqG1}, \ref{eq:case2g2asymp})
exhibit aging, since they depend on the age of the process $t$. 

Considering the asymptotic behavior of $C(t,t')$ for large \emph{t},\[
\hat{C}(t,u)\approx\]
\begin{equation}
\frac{1}{u}\frac{\tau_{+}}{A_{-}\Gamma(\alpha_{-})t^{1-\alpha_{-}}}\left\{ 1-\frac{\left[1-\hat{\psi}_{+}(u)\right]\left[1-\hat{\psi}_{-}(u)\right]}{\tau_{+}u\left[1-\hat{\psi}_{-}(u)\hat{\psi}_{+}(u)\right]}\right\} .\label{eqG11}\end{equation}
 This equation is similar to Eq. (\ref{eqVerO}), especially if we
notice that the {}``effective mean'' time of state \emph{off} until
total time \emph{t} scales as $A_{-}t^{1-\alpha_{-}}$. 

For the special case, where \emph{on} times are exponentially distributed,
the correlation function \emph{C} is a product of two identical expressions
\emph{for all} \emph{t} and $t'$: 

\begin{equation}
\hat{C}(s,u)=\frac{\tau_{+}}{1+s\tau_{+}-\psi_{-}(s)}\cdot\frac{\tau_{+}}{1+u\tau_{+}-\psi_{-}(u)},\label{eqtwoid}\end{equation}
 where \emph{s} (\emph{u}) is the Laplace conjugate of \emph{t} ($t'$)
respectively. Comparing to Eq. (\ref{eq:Iexpon}) we obtain \begin{equation}
C(t,t')=\langle I(t)\rangle\langle I(t')\rangle,\label{eqcprod}\end{equation}
 and for the normalized correlation function \begin{equation}
g^{(2)}(t,t')=\frac{\left\langle I(t')\right\rangle }{\left\langle I(t+t')\right\rangle }.\label{eqg2ratio}\end{equation}
 Eqs. (\ref{eqg2ratio}, \ref{eqcprod}) are important since they
show that measurement of mean intensity $\langle I(t)\rangle$ yields
the correlation functions, for this case. While our derivation of
Eqs. (\ref{eqg2ratio}, \ref{eqcprod}) is based on the assumption
of exponential on times, it is valid more generally for any $\psi_{+}(t)$
with finite moments, in the asymptotic limit of large $t$ and $t'$.
To see this note that Eqs. (\ref{eqG1}, \ref{eqiii}) yield $C(t,t')\sim\langle I(t)\rangle\left\langle I(t')\right\rangle $.
\begin{figure}
\includegraphics[%
  width=1.0\columnwidth,
  keepaspectratio]{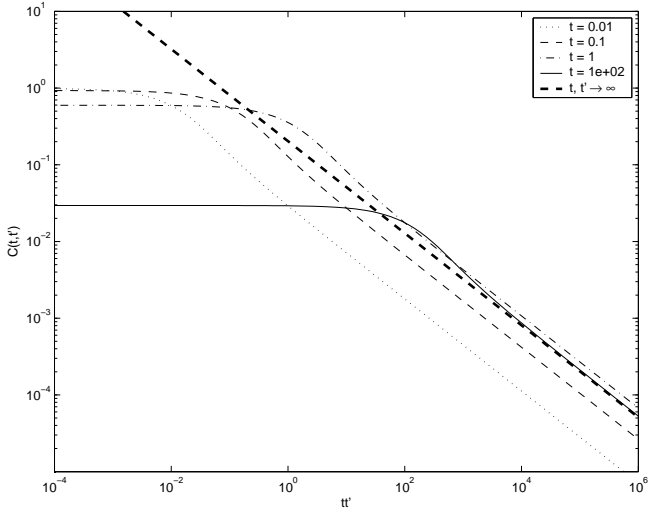}

\caption{\label{fig3} Exact $C(t,t')$ for Case $2$: exponential \emph{on}
times and power law \emph{off} times with $\alpha_{-}=0.4$. We use
$\hat{\psi}_{+}(s)=1/(1+s)$ and $\hat{\psi}_{-}(s)=1/(1+s^{0.4})$
and numerically obtain the correlation function. For each curve in
the figure we fix the time $t$. The process starts in the state $on$.
Thick dashed straight line shows the asymptotic behavior Eq. (\ref{eqG1}).
For short times ($t'<1$ for our example) we observe the behavior
$C(t,t')\sim C(t,0)=\langle I(t)\rangle$, the correlation function
is flat.}
\end{figure}

In Fig. \ref{fig3} we compare the asymptotic result (\ref{eqG1})
with exact numerical double Laplace inversion of the correlation function.
We use exponential PDF of \emph{on} times $\psi_{+}(s)=1/(1+s)$,
and power law distributed \emph{off} times: $\hat{\psi}_{-}(s)=\hat{\psi}_{-}(s)=1/(1+s^{0.4})$
corresponding to $\alpha_{-}=0.4$. Convergence to asymptotic behavior
is observed. 

\textbf{Remark} For fixed $t$ the correlation function in Eq. (\ref{eqG1})
exhibits a $(t')^{\alpha_{-}-1}$ decay. A $(t')^{\alpha_{-}-1}$
decay of an intensity correlation function was reported in experiments
of Orrit's group \cite{Verberk02} for uncapped NCs (for that case
$\alpha_{-}=0.65\pm0.2$). However, the measured correlation function
is a time averaged correlation function Eq. (\ref{eqTA}) obtained
from a single trajectory. In that case the correlation function is
independent of $t$, and hence no comparison between theory and experiment
can be made yet.

\subsection{Case 3}

We now consider case $3$, and find \cite{MB_JCP04}\begin{equation}
C(t,t')=P_{+}-P_{+}P_{-}\frac{\sin\pi\alpha}{\pi}B\left(\frac{1}{1+t/t'};1-\alpha,\alpha\right),\label{eq:case3asympcorr}\end{equation}
 where \[
P_{\pm}=\frac{A_{\pm}}{A_{+}+A_{-}}\]
 following from Eq. (\ref{eqIave}), and where \[
B(z;a,b)=\int_{0}^{z}x^{a-1}(1-x)^{b-1}dx\]
 is the incomplete beta function. The behavior in this limit does
not depend on the detailed shape of the PDFs of the \emph{on} and
\emph{off} times, besides the parameters $A_{+}/A_{-}$ and $\alpha$.
We note that both terms of Eq. (\ref{eq:Corr}) contribute to Eq.
(\ref{eq:case3asympcorr}). The appearance of the incomplete beta
function in Eq. (\ref{eq:case3asympcorr}) is related to the concept
of persistence. The probability of not switching from state \emph{on}
to state \emph{off} in a time interval $(t,t+t')$, assuming the process
is in state \emph{on} at time $t$, is called the persistence probability.
In the scaling limit this probability is\begin{equation}
P_{0}(t,t+t')\sim1-\frac{\sin\pi\alpha}{\pi}B\left(\frac{1}{1+t/t'};1-\alpha,\alpha\right).\label{eqPers}\end{equation}
 The persistence implies that long time intervals in which the process
does not jump between states \emph{on} and \emph{off}, control the
asymptotic behavior of the correlation function. The factor $P_{+}$,
which is controlled by the amplitude ratio $A_{+}/A_{-}$, determines
the expected short and long time $t'$ behaviors of the correlation
function, namely $C(\infty,0)=\lim_{t\rightarrow\infty}\langle I(t)I(t+0)\rangle=P_{+}$
and $C(\infty,\infty)=\lim_{t\rightarrow\infty}\langle I(t)I(t+\infty)\rangle=(P_{+})^{2}$.
With slightly more details the two limiting behaviors are: \begin{equation}
C(t,t')\sim\left\{ \begin{array}{cc}
P_{+} & \,\,\,\frac{t'}{t}\ll1\\
\,\, & \,\,\\
(P_{+})^{2}+P_{+}P_{-}\frac{\sin(\pi\alpha)}{\pi\alpha}\left(\frac{t'}{t}\right)^{-\alpha} & \,\,\,\frac{t'}{t}\gg1.\end{array}\right.\label{eq:case3asympasymp}\end{equation}
 Using Eq. (\ref{eqIave}) the normalized intensity correlation function
is $g^{(2)}(t,t')\sim C(t,t')/(P_{+})^{2}$. 

\begin{figure}
\includegraphics[%
  width=1.0\columnwidth,
  keepaspectratio]{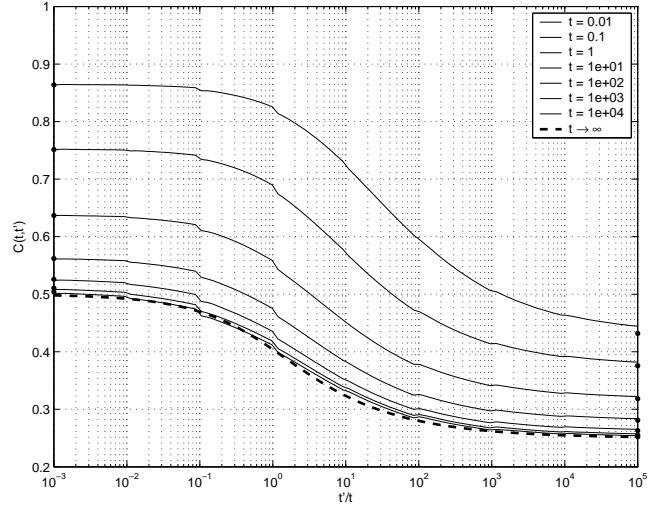}

\caption{\label{cap:alpha0.4exact}Exact $C(t,t')$ for case $3$, when both
\emph{on} and \emph{off} times are power law distributed with $\alpha=0.4$.
We use $\hat{\psi}_{\pm}(s)=1/(1+s^{0.4})$ for different times \emph{t}
increasing from the topmost to the lowermost curves. The dots on the
left and on the right show $C(t,0)=\left\langle I(t)\right\rangle $
and $C(t,\infty)=\left\langle I(t)\right\rangle /2$ respectively.
The process starts in the state \emph{on}.}
\end{figure}

In Fig. \ref{cap:alpha0.4exact} we compare the asymptotic result
(\ref{eq:case3asympcorr}) with exact numerical double Laplace inversion
of the correlation function for PDFs $\hat{\psi}_{+}(s)=\hat{\psi}_{-}(s)=1/(1+s^{0.4})$.
Convergence to Eq. (\ref{eq:case3asympcorr}) is seen. 

\textbf{Remark} For small $t'/t$ we get flat correlation functions.
Flat correlation functions were observed by Dahan's group \cite{Dahan}
for capped NCs. However, the measured correlation function is a single
trajectory correlation function Eq. (\ref{eqTA}), and hence no comparison
between theory and experiment can be made yet.

\section{Non Ergodicity}

Non-ergodicity of blinking quantum dots was first pointed out in the
experiments of the group of Dahan \cite{Dahan}. We begin the discussion
of nonergodicity in blinking NCs by plotting 100 time averaged correlation
functions from 100 NCs in Fig.%
\begin{figure}
\includegraphics[%
  width=1.0\columnwidth,
  keepaspectratio]{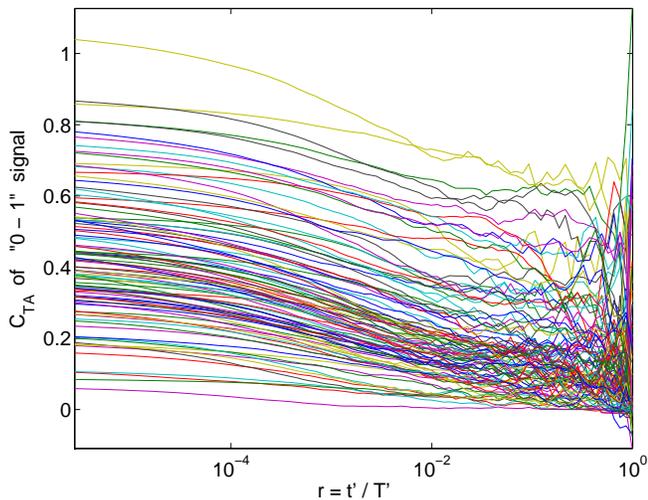}

\caption{\label{cap:100-traj}100 experimental time averaged correlation functions
(one of which is obtained from the signal shown in Fig. \ref{cap:Kuno1}),
after {}``renormalizing'' the average on and off intensities to
be 1 and 0, respectively. Note logarithmic abscissa.}
\end{figure}
\ref{cap:100-traj}. Clearly, correlation functions obtained are different.
The simplest explanation would be that the NCs have different statistical
properties. However, similar variability is also observed for a given
NC, when we calculate correlation functions for different $T'$ (e.g.,
\cite{Dahan}). To further illustrate this point, we generate on a
computer the two state process, with power law waiting time of on
and off times following $\psi_{-}(\tau)=\psi_{+}(\tau)=\alpha\tau^{-1-\alpha}$
for $\tau>1$ (and zero otherwise). For each trajectory we calculate
its own time average correlation. As we show in Fig.%
\begin{figure}
\includegraphics[%
  width=1.0\columnwidth,
  keepaspectratio]{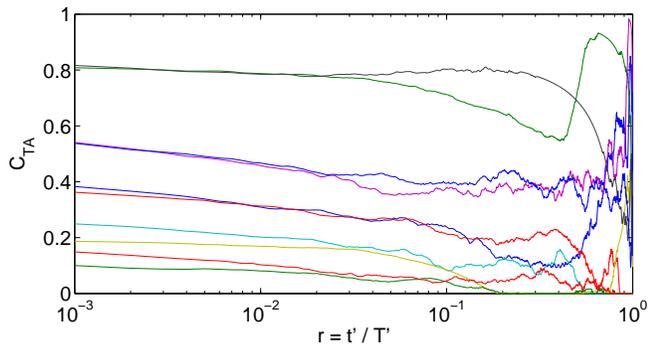}

\caption{\label{cap:Ctasimul0.8}Ten typical simulated realizations of $C_{TA}$
for $\alpha=0.8$.}
\end{figure}
\ref{cap:Ctasimul0.8} the trajectories exhibit ergodicity breaking.
The most striking feature of the figure is that even though trajectories
are statistically identical, the correlation function of the process
is random, similar to the experimental observation. In complete contrast,
if we consider a two state process with on and off times following
exponential statistics, then all the correlation functions would be
identical, and all of them would follow the same master curve: the
ensemble average correlation function. 

In this section we consider the non-ergodic properties of the blinking
NCs using a stochastic approach. We assume that all the NCs are statistically
identical in agreement with \cite{Brokmann03}, and restrict ourselves
to the Case 3. For the sake of simplicity we only consider the case
when distribution of \textit{on} times is identical to distribution
of \textit{off} times, namely $\alpha_{-}=\alpha_{+}=\alpha$ and
$A_{+}=A_{-}=A$. Generalization to $A_{+}\neq A_{-}$ is straightforward
\cite{BelBarkai}. The nonergodicity is found only for $\alpha<1$,
when the mean transition time is infinite, and should therefore disappear
when exponential cutoffs of off and on times become relevant \cite{ChungBawendi04},
i.e., when the mean transition times become of the order, or less
than the experimental time. The described model, however, is valid
in a wide time window spanning many orders of magnitude for the NCs,
and is relevant to other systems, as mentioned in Section \ref{sec:Physical-Meaning}.

\subsection{Distribution of time averaged intensity}

As mentioned in the introduction, the blinking NCs exhibit a non-ergodic
behavior. In particular the ensemble average intensity $\langle I\rangle$
is not equal to the time average $\overline{I}$. Of course in the
ergodic phase, namely when both the mean \textit{on} and \textit{off}
times are finite, we have $\langle I\rangle=\overline{I}$, in the
limit of long measurement time. More generally we may think about
$\overline{I}$ as a random function of time, which will vary from
one measurement to another. In the ergodic phase, and in the asymptotic
limit the distribution of $\overline{I}$ approaches a delta function
\begin{equation}
P(\overline{I})\rightarrow\delta(\overline{I}-\langle I\rangle).\end{equation}
 The theory of non-ergodic processes deals with the question what
is the distribution of $P(\overline{I})$ in the non-ergodic phase.
For the two state stochastic model \begin{equation}
\overline{I}=\frac{T^{+}}{T}\end{equation}
 where $T^{+}$ is the total time spent in state \textit{on}. 

A well known example of similar ergodicity breaking is regular diffusion,
or a binomial random walk on a line. The walker starts at the origin
and can go left or right randomly, at each step. Let the measurement
time be $t$, and the position of the random walker be $x(t)$. The
total time the walker remains on the right of the origin $x(t)>0$
is $T^{+}$. The PDF of return time (or of number of steps) $\tau$
to the origin decays as $\tau^{-3/2}$ for large $\tau$, so that
$\alpha=1/2$. Two half-axes at both sides of the origin can be thought
of as the two states, on and off, of the random walker. The well-established
result is that the fraction $\overline{I}$ of total time spent by
the walker on either side, in the long time limit is given by the
arcsine law \cite{Feller,GL}\[
P\left(\overline{I}\right)=\frac{1}{\pi\sqrt{\overline{I}(1-\overline{I})}}.\]
 A main feature of this PDF is its divergence at $\overline{I}=0,\,1$,
indicating that the random walker will most probably spend most of
its time on one side (either left or right) of the origin. In particular
the naive expectation that the particle will spend half of its time
on the right and half on the left, in the limit of long measurement
time, is wrong. In fact the minimum of the arcsine PDF is on $\overline{I}=\langle\overline{I}\rangle=1/2$.
In other words the ensemble average $\langle\overline{I}\rangle=1/2$
is the least likely event. This result might seem counter intuitive
at first, but it is due to the fact that the mean time for return
to the origin is infinite. This in turn means that the particle gets
randomly stuck on $x<0$ or on $x>0$ for a period which is of the
order of the measurement time, no matter how long this measurement
time is. 

In the more general case $0<\alpha<1$ the distribution of $\overline{I}$
can be calculated based on the work of Lamperti \cite{Lamperti58}
(see also \cite{GL}), and one finds \begin{equation}
l_{\alpha}(\overline{I})=\frac{\sin\left(\pi\alpha\right)}{\pi}\frac{\overline{I}^{\alpha-1}\left(1-\overline{I}\right)^{\alpha-1}}{\overline{I}^{2\alpha}+\left(1-\overline{I}\right)^{2\alpha}+2\cos\left(\pi\alpha\right)\overline{I}^{\alpha}\left(1-\overline{I}\right)^{\alpha}},\label{eqLAmp}\end{equation}
 which is shown in Fig.%
\begin{figure}
\includegraphics[%
  width=1.0\columnwidth,
  keepaspectratio]{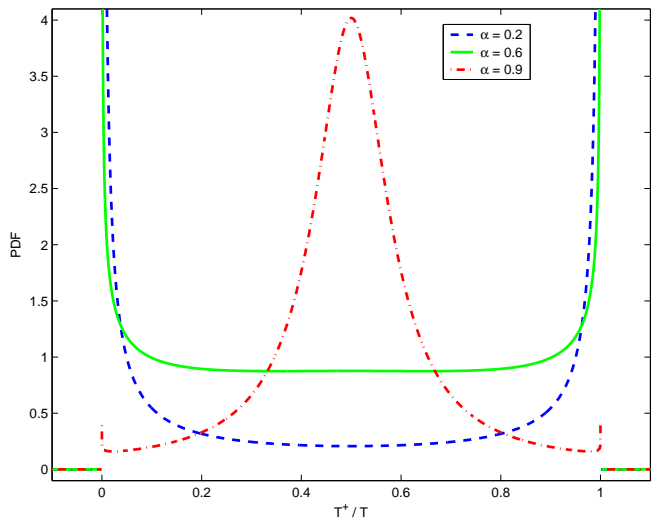}

\caption{The probability density function of $\overline{I}=T^{+}/T$ for the
case $\psi_{+}(t)=\psi_{-}(t)\propto t^{-(1+\alpha)}$. For the ergodic
phase $\alpha>1$ $P(\overline{I})$ is a delta function on $\langle I\rangle=1/2$.
In the non-ergodic phase $\overline{I}$ is a random function, for
small values of $\alpha$ the $P(\overline{I})$ is peaked on $\overline{I}=0$
and $\overline{I}=1$, indicating a trajectory which is in state \textit{off}
or \textit{on} for a period which is of the order of measurement time
$T$. }

\label{fig6}
\end{figure}
\ref{fig6}. When $\alpha\rightarrow0$ the PDF of $\overline{I}$
is peaked around $\overline{I}=0$ and $\overline{I}=1$, corresponding
to blinking trajectories which for most of the observation time $T$
are in state \emph{off} or state \emph{on} respectively. When $\alpha\rightarrow1$,
we see that $l_{\alpha}(\overline{I})$ attains a maximum when $\overline{I}=\langle I\rangle=1/2$,
indeed in the ergodic phase $\alpha>1$ we obtain as expected a delta
peak centered on $\overline{I}=1/2$, as we mentioned. There exists
a critical $\alpha_{c}=0.594611...$ above (under) which $l_{\alpha}(\overline{I})$
has a maximum (minimum) on $\overline{I}=1/2$. Note that the Lamperti
PDF in Eq. (\ref{eqLAmp}) is not sensitive to the precise shapes
of the \emph{on} and \emph{off} time distributions (besides $\alpha$
of course). For situations in which $A_{-}\ne A_{+}$ the symmetry
of the Lamperti PDF will not hold. Note that line shapes with structures
similar to those in Fig. \ref{fig6}, were obtained by Jung et al.
\cite{Jung02} in a related problem. Similar expressions are also
used in stochastic models of spin dynamics \cite{Bald}, and in general,
the problem of occupation times, and a related persistence concept,
are of a wide interest in different fields \cite{Dhar99,Majumdar,BelBarkai}. 

Next we extend our understanding of the \emph{distribution} of time
averaged intensity to the time averaged correlation functions defined
in Eq. (\ref{eqTA}).

\subsection{Distribution of time averaged correlation function}

\begin{figure}
\includegraphics[%
  width=1.0\columnwidth,
  keepaspectratio]{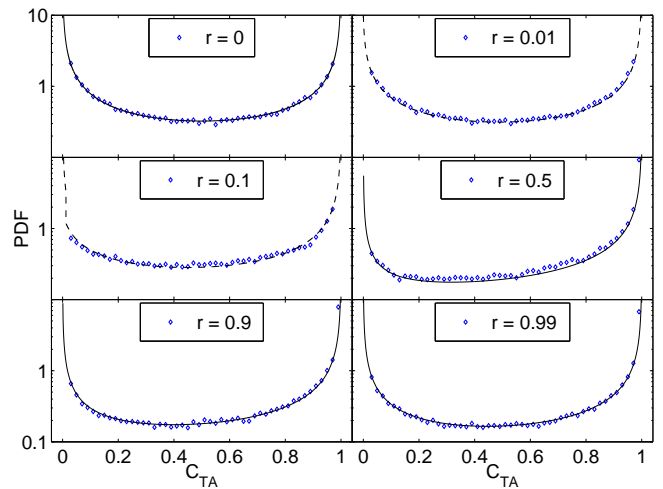}

\caption{\label{cap:theta0.3}PDF of $C_{TA}(T',t')$ for different $r=t'/T'$
and $\alpha=0.3$. Abscissas are possible values of $C_{TA}(T',t')$.
Diamonds are numerical simulations. Curves are analytical results
without fitting: for $r=0$ Eq. (\ref{eqLAmp}) is used (full line),
for $r=0.01$ and 0.1 Eq. (\ref{eq12}) is used (dashed) and for $r=0.5$,
0.9 and 0.99 Eq. (\ref{eq17}) is used (full).}
\end{figure}
\begin{figure}
\includegraphics[%
  width=1.0\columnwidth,
  keepaspectratio]{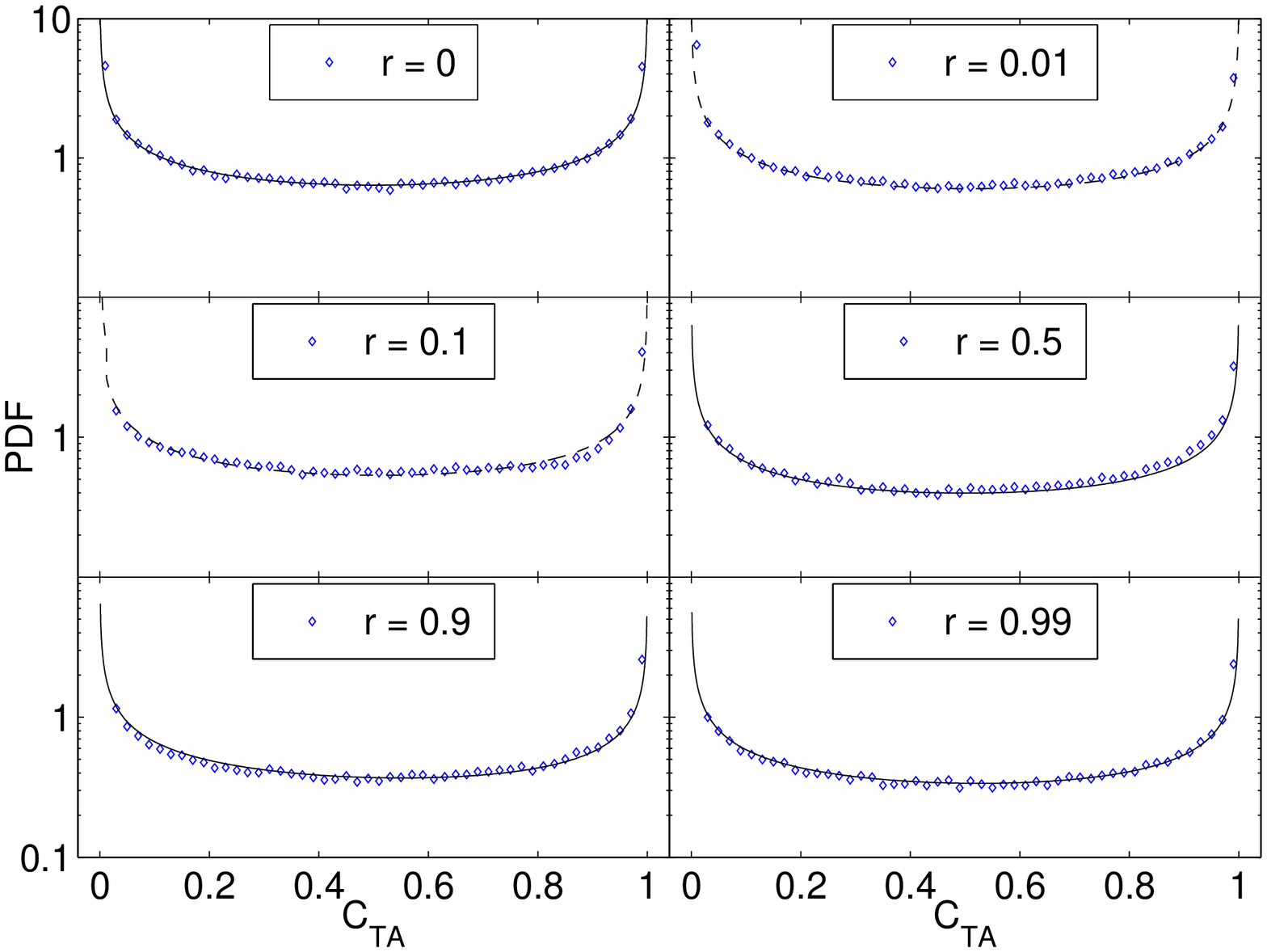}

\caption{\label{cap:theta0.5}PDF of $C_{TA}(T',t')$ for different $r=t'/T'$
and $\alpha=0.5$. Diamonds are numerical simulations. Curves are
analytical results without fitting: for $r=0$ Eq. (\ref{eqLAmp})
is used (full line), for $r=0.01$ and 0.1 Eq. (\ref{eq12}) is used
(dashed) and for $r=0.5$, 0.9 and 0.99 Eq. (\ref{eq17}) is used
(full).}
\end{figure}
\begin{figure}
\includegraphics[%
  width=1.0\columnwidth,
  keepaspectratio]{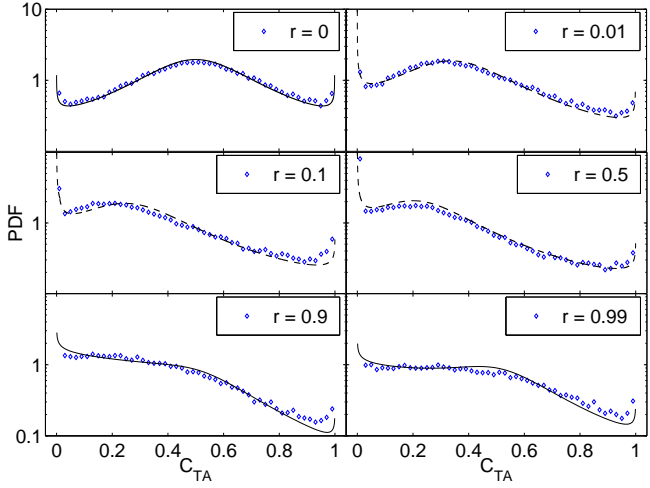}

\caption{\label{cap:theta0.8}PDF of $C_{TA}(T',t')$ for different $r=t'/T'$
and $\alpha=0.8$. Diamonds are numerical simulations. Curves are
analytical results without fitting: for $r=0$ Eq. (\ref{eqLAmp})
is used (full line), for $r=0.01$, 0.1 and 0.5 Eq. (\ref{eq12})
is used (dashed) and for $r=0.9$ and 0.99 Eq. (\ref{eq17}) is used
(full). If compared with the cases $\alpha=0.3$ and 0.5, the distribution
function exhibits a weaker non-ergodic behavior, namely for $r=0$
the distribution function peaks on the ensemble average value of $1/2$.}
\end{figure}

We first consider the non-ergodic properties of the correlation function
for the case $t'=0$. It is useful to define \begin{equation}
\mathcal{I}_{[a,b]}=\int_{a}^{b}I(t)\textrm{d}t/(b-a),\label{eq:Iab}\end{equation}
 the time average intensity between time $a$ and time $b>a$, and
\[
T=T'-t',\]
 \[
r=\frac{t'}{T'}.\]
 Using Eq. (\ref{eqTA}) and for $t'=0$ the time averaged correlation
function is identical to the time average intensity \begin{equation}
C_{TA}(T,0)=\mathcal{I}_{[0,T]}=\frac{T^{+}}{T},\label{eq07}\end{equation}
 and its PDF is given by Eq. (\ref{eqLAmp}). Figs. \ref{cap:theta0.3},
\ref{cap:theta0.5} and \ref{cap:theta0.8} for the case $r=0$, show
these distributions for $\alpha=0.3$, 0.5 and 0.8, respectively,
together with the numerical results. 

An analytical approach to estimate the distributions $P_{C_{TA}(T',t')}(z)$
of $C_{TA}(T',t')=z$ for nonzero $t'$ was developed in \cite{MB_condmat05,MB_PRL05}.
To treat the problem a non-ergodic mean field approximation was used,
in which various time averages were replaced by the time average intensity
$\mathcal{I}_{[0,T]}$, \emph{specific for a given realization}. For
short $t'\ll T'$ the result is\begin{widetext}
\begin{equation}
C_{TA}(T',t')\simeq\left\{
\begin{array}{l l}
 {\mathcal{I}}_{[0,T]} \left\{ 1  - \left(1 - {\mathcal{I}}_{[0,T]} \right)
\left[ \left( {r \over {(1-r)\mathcal{I}}_{[0,T]} } \right)^{1 - \alpha} \left( { \sin \pi \alpha \over \pi \alpha} + 1 \right) - { \sin \pi \alpha \over \pi \alpha}{ r \over {(1-r)\mathcal{I}}_{[0,T]} } \right] \right\} 
&\ t' < T^{+} \\
{\mathcal{I}}_{[0,T]} ^2 &\ t' > T^{+}.
\end{array}
\right.
\label{eq12}
\end{equation}
\end{widetext} Eq. (\ref{eq12}) yields the correlation function, however unlike
standard ergodic theories the correlation function here is a random
function since it depends on $\mathcal{I}_{[0,T]}$. The distribution
of $C_{TA}(T',t')$ is now easy to find using the chain rule, and
Eqs. (\ref{eqLAmp},\ref{eq07}, \ref{eq12}). In Figs. \ref{cap:theta0.3},
\ref{cap:theta0.5} and \ref{cap:theta0.8} we plot the PDF of $C_{TA}(T',t')$
(dashed curves) together with numerical simulations (diamonds) and
find excellent agreement between theory and simulation, for the cases
where our approximations are expected to hold $r<1/2$. We observe
that unlike the $r=0$ case the PDF of the correlation function exhibit
a non-symmetrical shape. To understand this note that trajectories
with short but finite total time in state \emph{on} ($T^{+}\ll T$)
will have finite correlation functions when $t'=0$. However when
$t'$ is increased the corresponding correlation functions will typically
decay very fast to zero. On the other hand, correlation functions
of trajectories with $T^{+}\sim T$ don't change much when $t'$ is
increased (as long as $t'\ll T^{+}$). This leads to the gradual nonuniform
shift to the left, and {}``absorption'' into $C_{TA}(T',t')=0$,
of the Lamperti distribution shape, and hence to non-symmetrical shape
of the PDFs of the correlation function whenever $r\ne0$.

We now turn to the case $T\ll t'$. Then\begin{equation}
C_{TA}(T',t')\simeq\mathcal{I}_{[0,T]}\mathcal{I}_{[t',T']}.\label{eq13}\end{equation}
 In the limit $t'/T'\rightarrow1$ this yields\begin{equation}
P_{C_{TA}(T',t')}(z)\sim[\ell_{\alpha}(z)+\delta(z)]/2,\label{eq:halflamperti}\end{equation}
 which is easily understood if one realizes that in this limit $\mathcal{I}_{[t',T']}$
in Eq. (\ref{eq13}) is either 0 or 1 with probabilities 1/2, and
that the PDF of $\mathcal{I}_{[0,T]}$ is Lamperti's PDF Eq. (\ref{eqLAmp}).
More generally, using the Lamperti distribution for $\mathcal{I}_{[0,T]}$,
and probabilistic arguments \cite{MB_condmat05}, the PDF of $C_{TA}(T',t')$
is approximated by\begin{widetext}
\begin{equation}
\begin{array}{l l}
 P_{ C_{TA}(T',t') } \left( z \right)  \simeq 
\left[ 1 - P_0\left(T, T' \right) \right] 
\left\{ \left[ 1 - P_0\left( t' , T' \right) \right] \int_z^1 { l_{\alpha} \left( x \right) \over x} {\textrm{d}} x + 
{ P_0\left( t',T' \right) \over 2} \left[ l_{\alpha} \left( z \right) + \delta\left( z \right) \right] \right\} 
+ P_0\left( T, T' \right) \left[ z l_{\alpha}\left( z \right) + { \delta\left( z \right) \over 2} \right],
\end{array}
\label{eq17}
\end{equation}
\end{widetext}where $P_{0}(a,b)$ is the persistence probability Eq. (\ref{eqPers}).
Note that to derive Eq. (\ref{eq17}) we used the fact that $\mathcal{I}_{[0,T]}$
and $\mathcal{I}_{[t',T']}$ are correlated. In Figs. \ref{cap:theta0.3},
\ref{cap:theta0.5} and \ref{cap:theta0.8} we plot these PDFs of
$C_{TA}(T',t')$ (solid curves) together with numerical simulations
(diamonds) and find good agreement between theory and simulation,
for the cases where these approximations are expected to hold, $r>1/2$.
In the limit $t'/T'\rightarrow1$ Eq. (\ref{eq17}) simplifies to
Eq. (\ref{eq:halflamperti}).

\section{Experimental evidence\label{sec:Experimental-evidence}}

\begin{figure}
\begin{tabular}{c}
\includegraphics[%
  width=1.0\columnwidth,
  keepaspectratio]{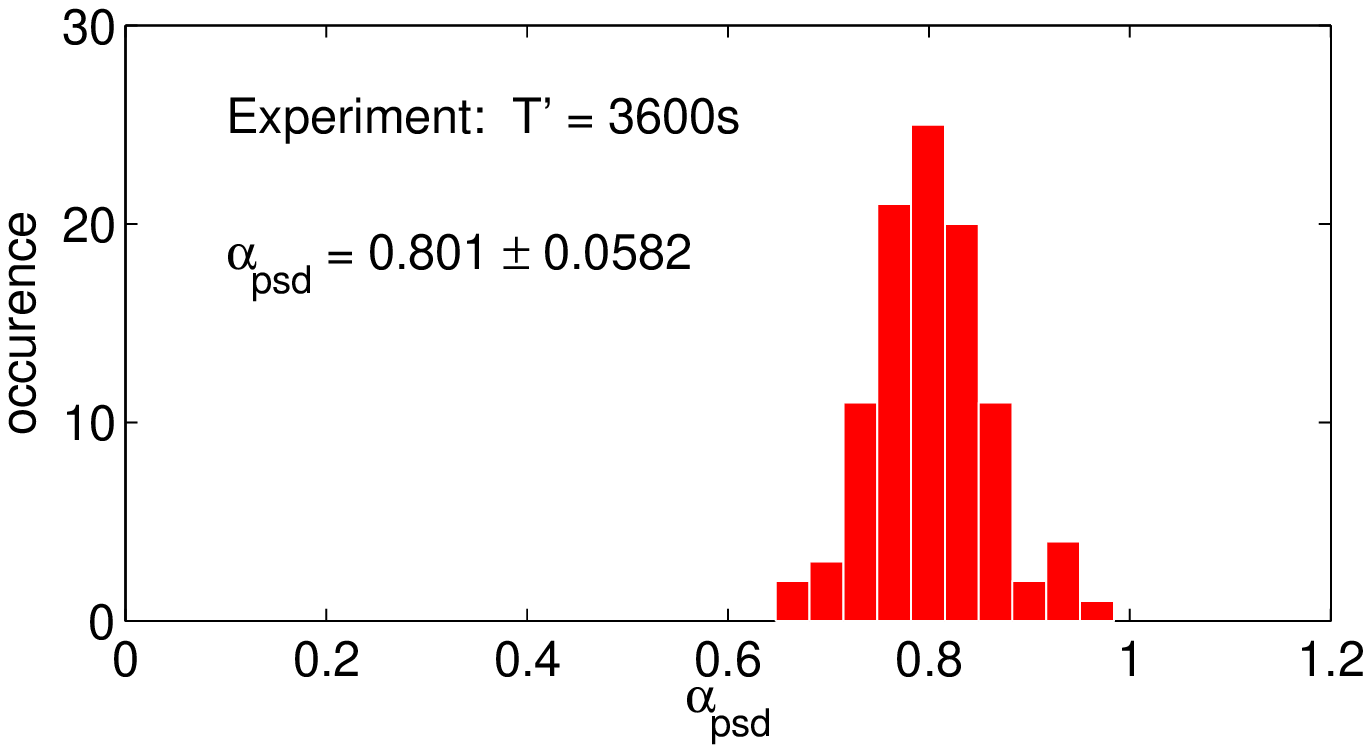}\tabularnewline
\includegraphics[%
  width=1.0\columnwidth,
  keepaspectratio]{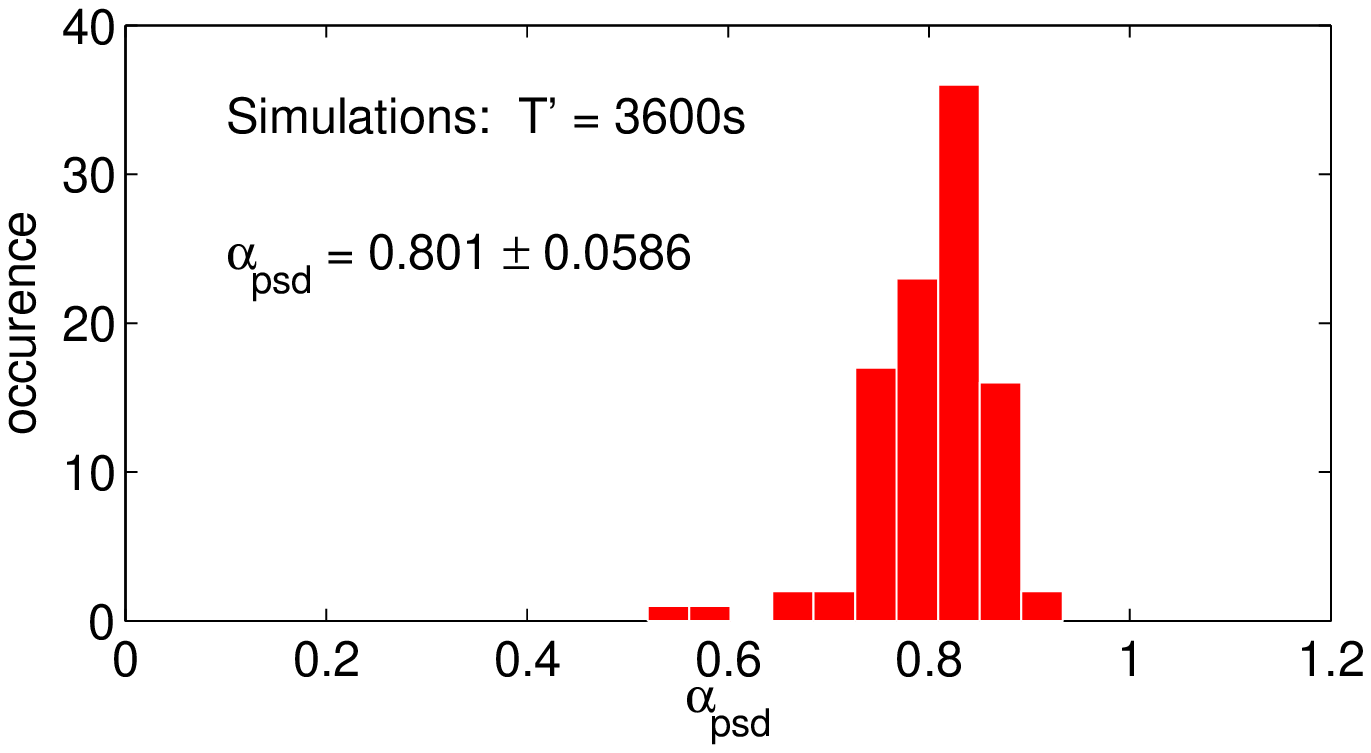}\tabularnewline
\end{tabular}

\caption{\label{cap:Histograms-of-thetapsd}Histograms of experimental (top)
and simulated (bottom) fitted values of $\alpha$ for 100 trajectories.
Fits are made to the power spectral densities of individual trajectories.}
\end{figure}

In this section we analyze experimental data and make comparisons
with theory. Data was obtained for 100 CdSe-ZnS nanocrystals at room
temperature %
\footnote{Core radius 2.7 nm with less than 10\% dispersion, 3 monolayers of
ZnS, covered by mixture of TOPO, TOP and TDPA. Quantum dots were spin
coated on a flamed fused silica substrate. CW excitation at 488 nm
of Ar$^{\textrm{+}}$ laser was used, excitation intensity in the
focus of oil immersion objective ($\textrm{NA}=1.45$) was $\sim600\textrm{W}/\textrm{cm}^{2}$.%
}. We first performed data analysis (similar to standard approach)
based on distribution of on and off times and found that $\alpha_{+}=0.735\pm0.167$
and $\alpha_{-}=0.770\pm0.106$ %
\footnote{The standard deviation figures for $\alpha_{\pm}$ here and for $\alpha_{\textrm{psd}}$
in Fig. \ref{cap:Histograms-of-thetapsd} represent the standard deviations
of the distributions of the corresponding exponents, and not the errors
in determination of their mean value. We also note that the on time
distributions are less close to the power law decays than the off
times, partly due to the exponential cutoffs, and partly due to varying
intensities in the on state (cf. Figs. \ref{cap:Kuno1}, \ref{cap:Kuno2}).%
}, for the total duration time $T'=T=3600$s (bin size 10ms, threshold
was taken as $0.16\max I(t)$ for each trajectory). Within error of
measurement, $\alpha_{+}\approx\alpha_{-}\approx0.75$. The value
of $\alpha\approx0.75$ implies that simple diffusion model with $\alpha=0.5$
is not valid in this case. An important issue is whether the exponents
vary from one NC to another. In Fig.%
\ref{cap:Histograms-of-thetapsd} (top) we show distribution of $\alpha$
obtained from data analysis of power spectra. The power spectrum method
\cite{MB_condmat05} yields a single exponent $\alpha_{\textrm{psd}}$
for each stochastic trajectory (which is in our case $\alpha_{+}\approx\alpha_{-}\approx\alpha_{\textrm{psd}}$).
Fig. \ref{cap:Histograms-of-thetapsd} illustrates that the spread
of $\alpha$ in the interval $0<\alpha<1$ is not large. Numerical
simulation of 100 trajectories switching between 1 and 0, with $\psi_{+}(\tau)=\psi_{-}(\tau)$
and $\alpha=0.8$, and with the same number of bins as the experimental
trajectories, was performed and distribution of $\alpha$ values estimated
from power spectra is also shown in Fig. \ref{cap:Histograms-of-thetapsd}
(bottom). We observe some spread of measured values of $\alpha$,
which is similar to experimental behavior. This indicates that experimental
data is compatible with the assumption that all dots are statistically
identical (in our sample), in agreement with \cite{Brokmann03,Issac05}.

We also tested our nonergodic theory and calculated distribution of
relative on times $T^{+}/T$, i.e., of the ratios of the total time
in the state on to the total measurement time. These relative on times
are equivalent to the experimental time averaged intensities after
their {}``renormalization'' in a way making average intensity in
state on/off to be 1/0, respectively, in analogy to our model stochastic
process. Experimental and simulated distributions shown in Fig.%
\begin{figure*}
\begin{tabular}{cc}
\includegraphics[%
  width=1.0\columnwidth,
  keepaspectratio]{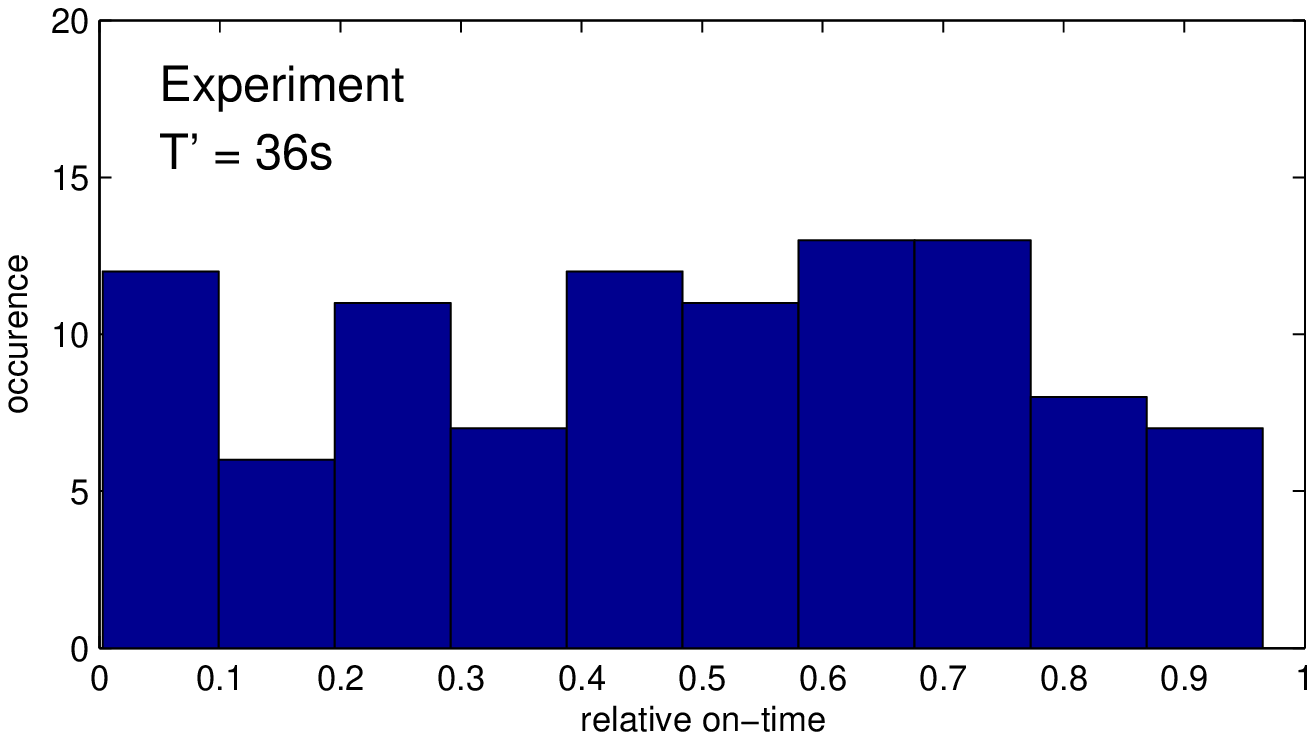}&
\includegraphics[%
  width=1.0\columnwidth,
  keepaspectratio]{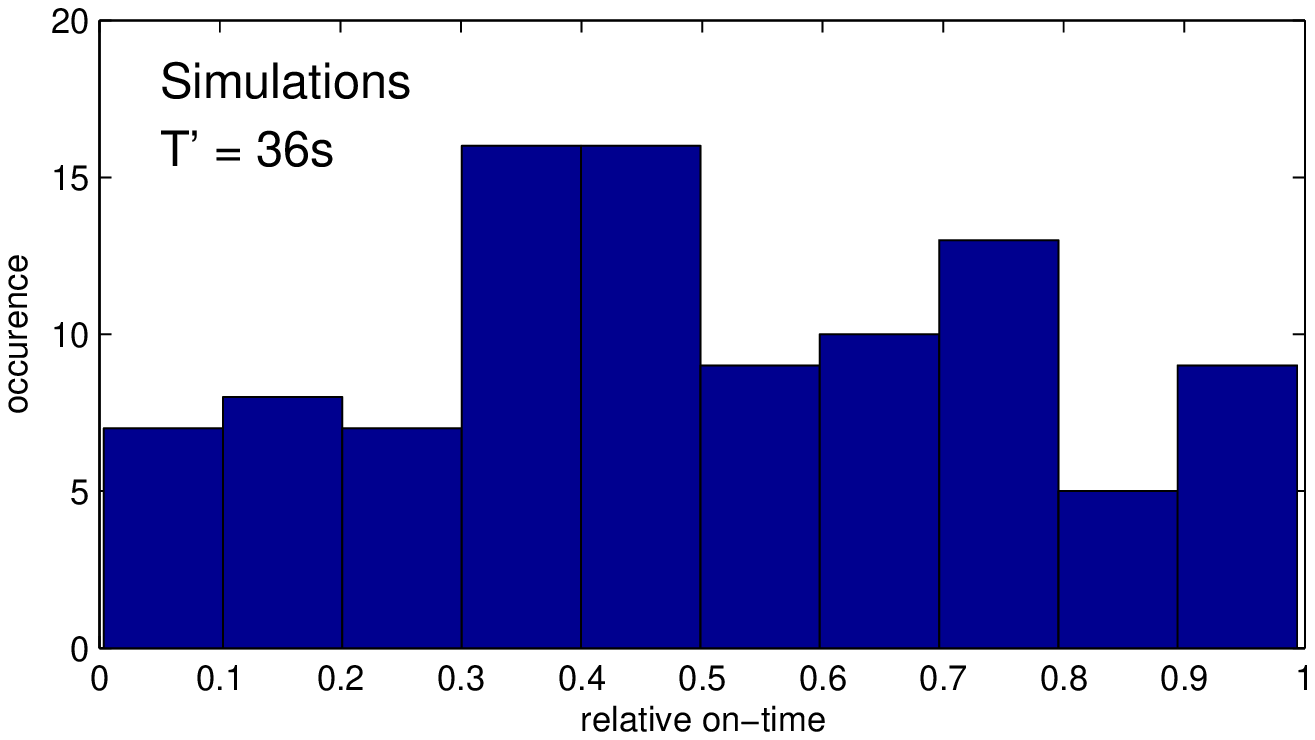}\tabularnewline
\includegraphics[%
  width=1.0\columnwidth,
  keepaspectratio]{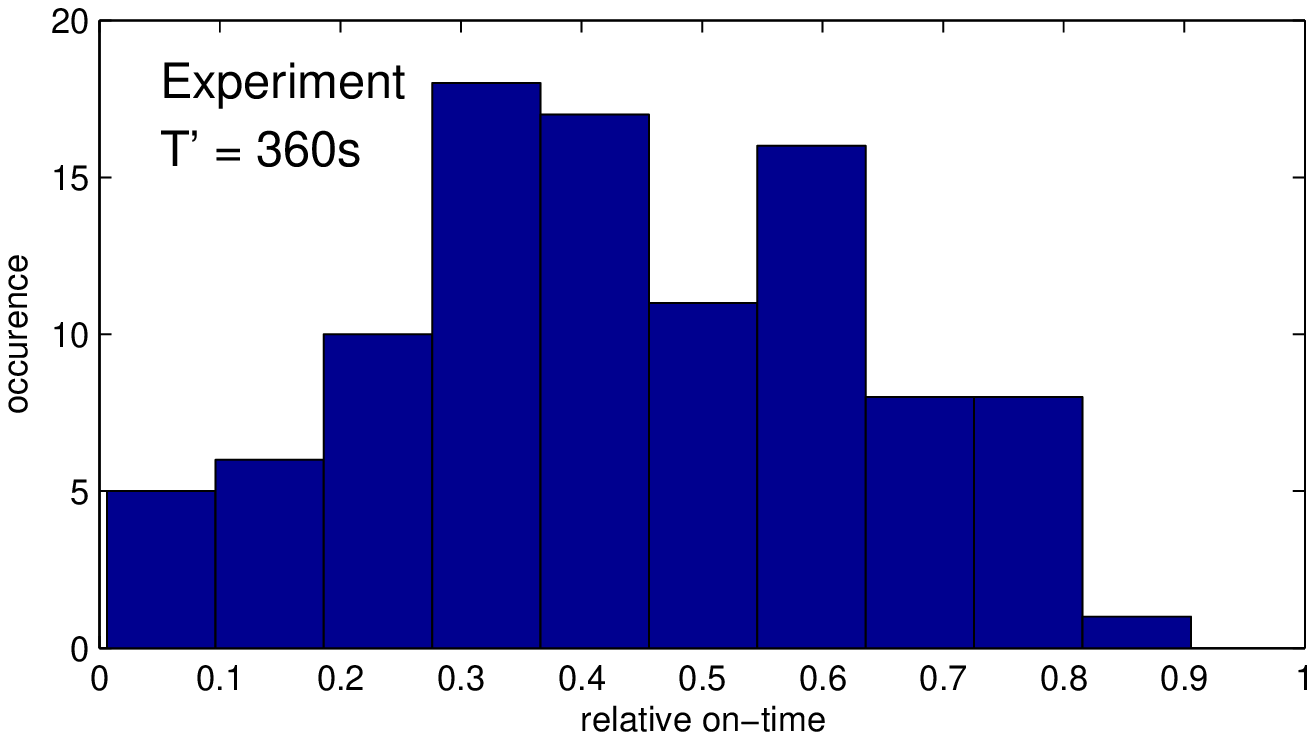}&
\includegraphics[%
  width=1.0\columnwidth,
  keepaspectratio]{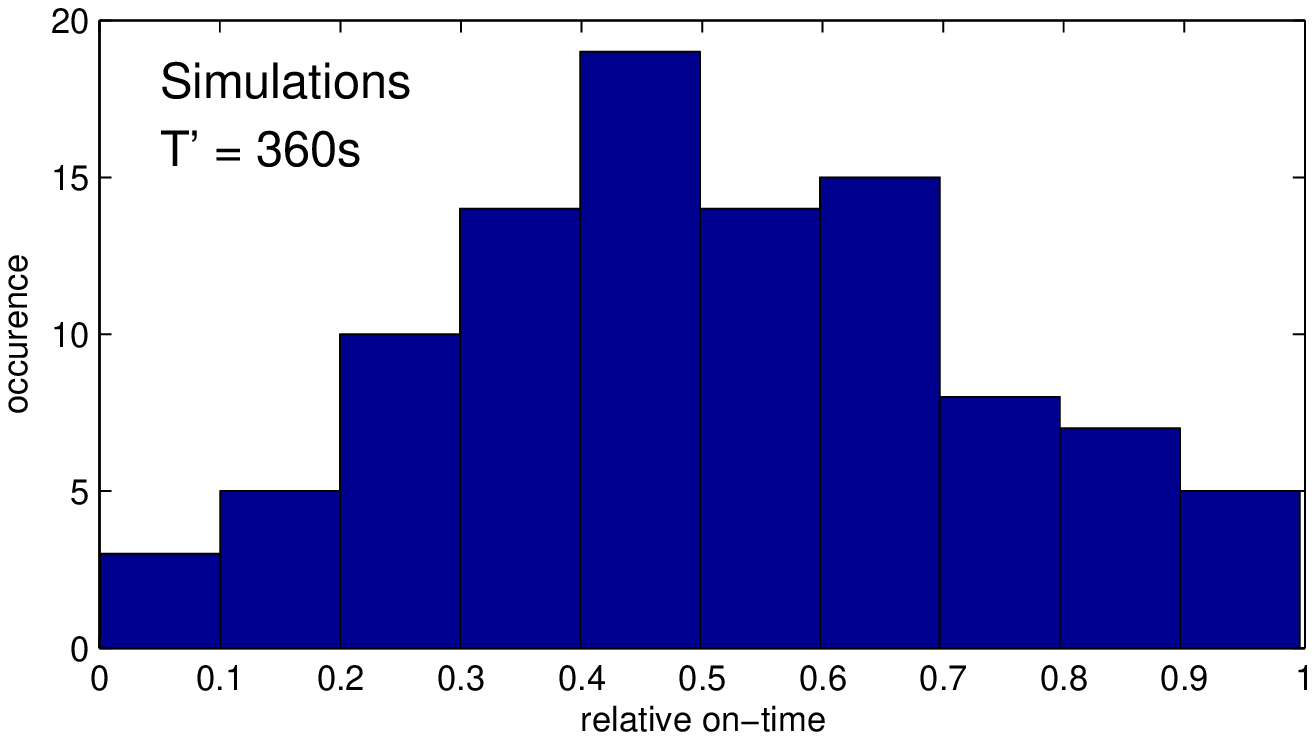}\tabularnewline
\includegraphics[%
  width=1.0\columnwidth,
  keepaspectratio]{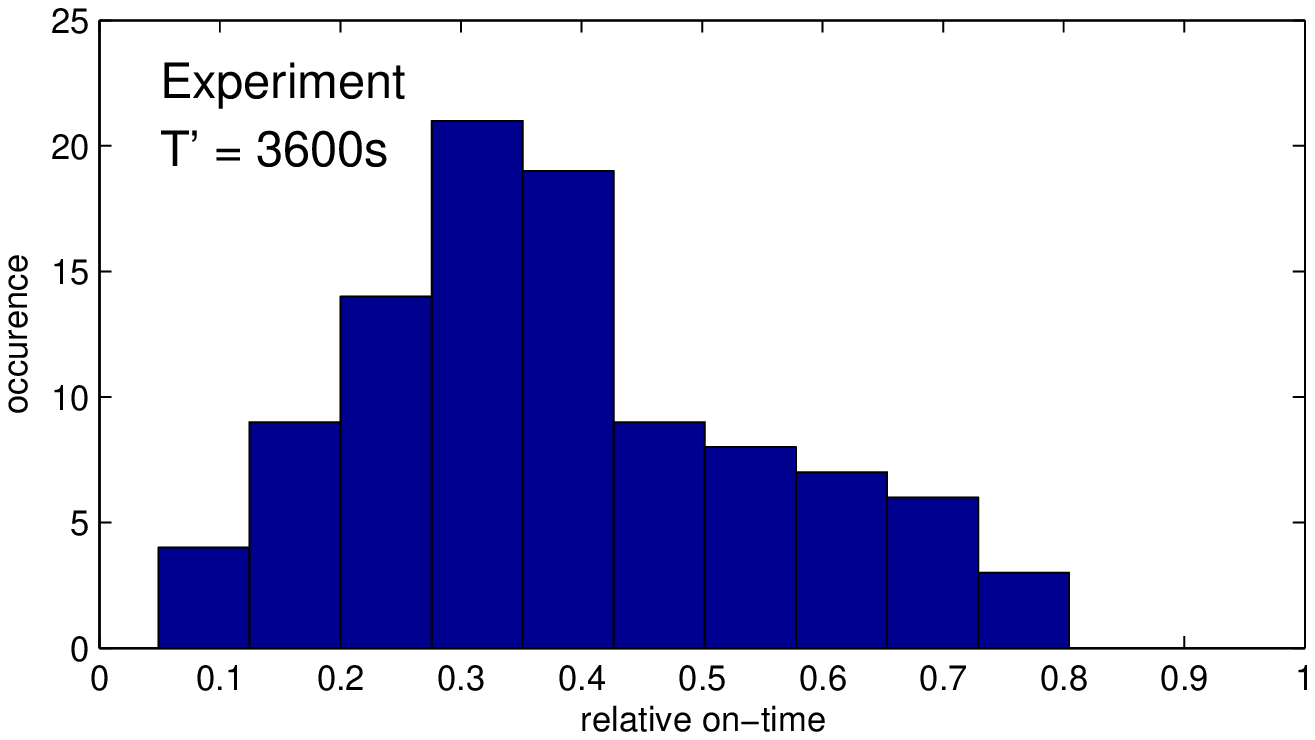}&
\includegraphics[%
  width=1.0\columnwidth,
  keepaspectratio]{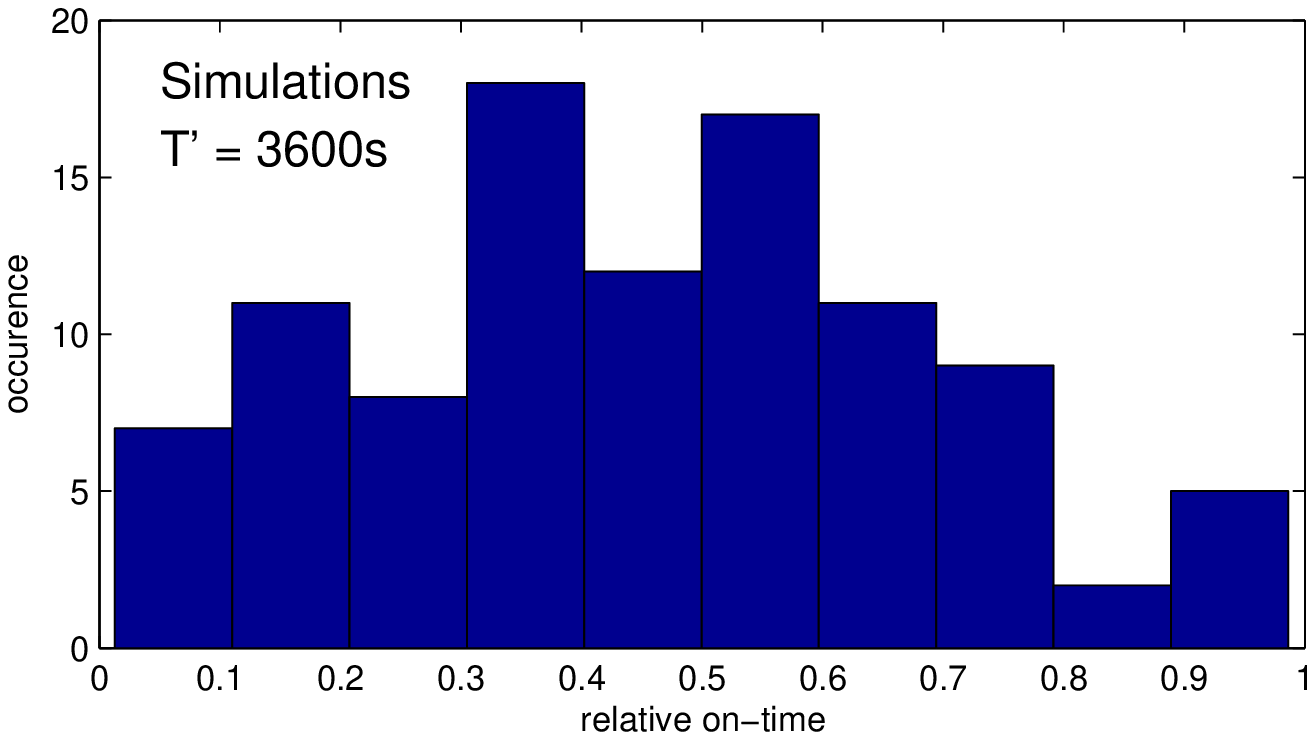}\tabularnewline
\end{tabular}

\caption{\label{cap:Histograms-of-ontime}Histograms of relative on times
$T^{+}/T$ for 100 experimental (left) and 100 simulated (right) intensity
trajectories, for different $T'$.}
\end{figure*}
\ref{cap:Histograms-of-ontime} are, overall, in good agreement. Two
important conclusions are derived from these distributions of relative
on times. First the data clearly exhibits ergodicity breaking: distribution
of relative on times is not delta peaked, instead it is wide in the
interval between 0 and 1, for different $T'$. The second important
conclusion is that for a reasonably chosen threshold (cf. Fig. \ref{cap:Kuno1}),
the experimental data is compatible with the assumption \[
\psi_{+}(\tau)\approx\psi_{-}(\tau),\]
 at least for a wide time window relevant to the experiments. In other
words, not only $\alpha_{+}\approx\alpha_{-}$ (ignoring the cutoffs)
but also $A_{+}\approx A_{-}$. This observation cannot be obtained
directly from the on and off time histograms like Fig. \ref{cap:Kuno2}
because if only power law tails are seen, as in Fig. \ref{cap:Kuno2},
these histograms cannot be normalized. To see that $A_{+}\approx A_{-}$
note that the distributions of relative on times are roughly symmetric
with respect to the median value of $1/2$ (cf. Fig. \ref{cap:Histograms-of-ontime}),
and the ensemble average of relative on times is also close to $1/2$,
while in general the ensemble average in our model process is given
by $A_{+}/(A_{+}+A_{-})$. In addition, the variance of the experimental
distributions for different $T'$ is close to the variance of the
Lamperti distribution $(1-\alpha)/4$ \cite{MB_condmat05} for $\alpha\approx0.8$.
There are a few comments to make. First, 100 trajectories are insufficient
to produce accurate histograms, as can be seen from the right side
of Fig. \ref{cap:Histograms-of-ontime}: ideally, these histograms
should be identical for different $T'$, and given by the Lamperti
distribution Eq. (\ref{eqLAmp}). Second, there is an effect due to
the signal discretization, leading to a flatter and wider histogram
at $T'=36$s. Third, there is a certain slow narrowing of the experimental
histogram as $T'$ increases, and the average relative on time slowly
decreases. Both of these trends are probably due to cutoffs in the
power law distributions, especially for on times, as can be seen in
Fig. \ref{cap:Kuno2}. These trends slightly depend on the choice
of the threshold separating on and off states.

As mentioned previously, the groups of Dahan and Bawendi \cite{Brokmann03,Shimizu01}
measure values of $\alpha_{+}\approx\alpha_{-}\approx0.5$ for hundreds
of quantum dots (see Table \ref{cap:Summary-on-off}), while we report
on a higher value of $\alpha$. An important difference between our
samples and Dahan/Bawendi groups is that in those works the dots are
embedded in PMMA, while in our case they are not {[}49{]} (see also
\cite{Issac05}).

\section{Summary and Conclusions}

Our main points are the following:

1. Simple three-dimensional diffusion model can be used to explain
the exponent $\alpha=1/2$ observed in many experiments. In some cases
deviations from $\alpha=1/2$ are observed, and modifications of Onsager
theory are needed. We cannot exclude other models.

2. Simple model of diffusion may lead to ergodicity breaking. Thus
ergodicity breaking in single molecule spectroscopy should not be
considered exotic or strange.

3. The time average correlation function is random. Ensemble average
correlation function exhibits aging. Hence data analysis should be
made with care.

4. Our data analysis shows $A_{+}\approx A_{-}$, $\alpha_{+}\approx\alpha_{-}$
(before the possible cutoffs) and that the distribution of $\alpha$
is narrow. It is important to check the validity of this result in
other samples of nanocrystals, since so far the main focus of experimentalist
was on values of $\alpha$ and not on the ratio of amplitudes $A_{+}/A_{-}$.

How general are our results? From a stochastic point of view ergodicity
breaking, L\'{e}vy statistics, anomalous diffusion, aging, and fractional
calculus, are all related. In particular ergodicity breaking is found
in other models with power law distributions, related to the underlying
stochastic model (the L\'{e}vy walk). For example the CTRW model
also exhibits ergodicity breaking \cite{BelBarkai}, and hence a natural
conflict with standard Boltzmann statistics emerges. Since power law
distributions are very common in natural behavior, we expect that
single particle ergodicity breaking will be a common theme. Further,
since we showed that a simple diffusion model can generate ergodicity
breaking, for the nano-crystals, we expect that ergodicity breaking
be found in other single molecule systems. One simple conclusion is
that predictions cannot be made, based on ensemble averages. In fact
the time averages of physical observables remain random even in the
limit of long measurement time. The fact that the time averaged correlation
function is a random function, means that some of the experimental
published results, on time average correlation functions, are not
reproducible.

\begin{acknowledgments}
Acknowledgment is made to the National Science Foundation for support
of this research with award CHE-0344930. EB thanks Center of Complexity,
Jerusalem.
\end{acknowledgments}
\bibliographystyle{apsrev}
\bibliography{NCs,KPL}

\end{document}